\documentclass[sigconf]{acmart}

\AtBeginDocument{%
  }

\copyrightyear{2025}
\acmYear{2025}
\setcopyright{acmlicensed}\acmConference[WWW Companion '25]{Companion
Proceedings of the ACM Web Conference 2025}{April 28-May 2, 2025}{Sydney,
NSW, Australia}
\acmBooktitle{Companion Proceedings of the ACM Web Conference 2025 (WWW
Companion '25), April 28-May 2, 2025, Sydney, NSW, Australia}
\acmDOI{10.1145/3701716.3715556}
\acmISBN{979-8-4007-1331-6/2025/04}

\usepackage{epigraph}
\begin{document}

% \title[Faces Speak Louder Than Words: Emotions Versus Textual Sentiment in the 2024 USA Presidential Election]{Faces Speak Louder Than Words: Emotions Versus Textual Sentiment in the 2024 USA Presidential Election}

\title[Faces Speak Louder Than Words: Emotions Versus Textual Sentiment \\ in the 2024 USA Presidential Election]{Faces Speak Louder Than Words: Emotions Versus Textual
Sentiment in the 2024 USA Presidential Election}


\author{Chiyu Wei}
\affiliation{%
  \institution{Department of Mathematics, \\ Dartmouth College}
  \city{Hanover}
  \state{New Hampshire}
  \country{USA}}
\email{chiyu.wei.gr@dartmouth.edu}

\author{Sean Noh}
\affiliation{%
  \institution{Quantitative Social Science, \\ Dartmouth College}
  \city{Hanover}
  \state{New Hampshire}
  \country{USA}}
\email{Sean.Noh.28@dartmouth.edu}

\author{Ho-Chun Herbert Chang}
\affiliation{%
  \institution{Quantitative Social Science, \\ Dartmouth College}
  \city{Hanover}
  \state{New Hampshire}
  \country{USA}}
\email{herbert.chang@dartmouth.edu}

\renewcommand{\shortauthors}{Chiyu Wei, Sean Noh, \& Ho-Chun Herbert Chang}
\begin{abstract}
Sentiment analysis of textual content has become a well-established solution for analyzing social media data. However, with the rise of images and videos as primary modes of expression, more information on social media is conveyed visually. Among these, facial expressions serve as one of the most direct indicators of emotional content in images. This study analyzes a dataset of Instagram posts related to the 2024 U.S. presidential election, spanning April 5, 2024, to August 9, 2024, to compare the relationship between textual and facial sentiment. Our findings reveal that facial expressions align with text sentiment, where positive sentiment aligns with happiness, although neutral and negative facial expressions provide critical information beyond negative valence. Furthermore, during politically significant events such as Donald Trump's conviction and assassination attempt, posts depicting Trump showed a 12\% increase in negative sentiment. Crucially, Democrats use their opponent's fear to depict weakness, whereas Republicans use their candidate's anger to depict resilience. Our research highlights the potential of integrating facial expression analysis with textual sentiment analysis to uncover deeper insights into social media dynamics.
\end{abstract}



  

\maketitle

\section{Introduction}

% \epigraphhead[20]{%
  \epigraph{Happy families are all alike; every unhappy family is unhappy in its own way.}{Anna Karenina (1877), Leo Tolstoy}
  % }
  
Sentiment analysis, the computational process of detecting and extracting emotional content from text or other forms of data, has emerged as a central tool in social science and affective computing \cite{pang2008opinion,liu2012sentiment}. From gauging public opinion on social media to understanding emotional responses to policy changes, sentiment analysis allows researchers to quantify subjective states and behaviors at scale. It has proven particularly useful for analyzing vast amounts of textual data, such as online reviews, news articles, and social media posts, offering insights into public sentiment, emotional trends, and decision-making processes.

Despite its utility, sentiment analysis has its limitations. Traditional approaches often rely on linguistic cues, which can be highly context-dependent and ambiguous. For instance, sarcasm, cultural nuances, and variations in language usage can significantly undermine the accuracy of text-based sentiment classification \cite{tang2014learning}. Furthermore, these methods often capture \textit{expressed} sentiment rather than the underlying emotional states, creating a gap between what individuals say and what they feel. As such, sentiment analysis remains an indirect proxy for measuring affective experiences.

If the aim is to measure emotion in its most raw and direct form, approaches beyond textual data must be considered. Affective computing, which seeks to interpret and simulate human emotions using computational models, increasingly turns to physiological and behavioral signals as richer sources of emotional information \cite{picard1997affective}. Among these, facial emotion recognition offers a direct, real-time window into an individual’s emotional state. Unlike text, facial expressions reflect spontaneous, involuntary responses that are difficult to consciously control, making them a more immediate and less filtered measure of emotion \cite{ekman1978facial}.

At the core of facial emotion recognition lies the analysis of facial \textit{Action Units} (AUs), which are specific muscle movements defined within the Facial Action Coding System (FACS) introduced by Ekman and Friesen \cite{ekman1978manual}. Action Units serve as the building blocks of facial expressions, enabling researchers to decompose complex emotions into measurable components. For example, the combination of AU6 (cheek raiser) and AU12 (lip corner puller) corresponds to a genuine smile (a Duchenne smile), while other configurations might indicate anger, sadness, or surprise. The literature surrounding AUs has expanded significantly in recent years, with applications ranging from psychology and marketing to computer vision and human-computer interaction \cite{calvo2010affect}.

In this paper, our principal research question is --- \textbf{How do textual and facial sentiment relate to and diverge from each other}? Using 501,011 captions and images from Instagram, we assess the strengths and limitations of using facial emotion for sentiment analysis. We first assess the divergences between dictionary-based and transformer-based sentiment with facial emotions. We then demonstrate its viability in computational social scientific comparisons, using changes in Trump's expression before and after his assassination attempt. We have the following research questions:

\begin{itemize}
    \item What is the distribution of facial sentiment on social media?
    \item How does facial emotion data map onto textual sentiment?
    \item How do partisans frame presidential candidates differently after major political events?
\end{itemize}

\section{Methodology}

\subsection{Data Acquisition}

To collect Instagram data, we first generated a list of keywords and hashtags relating to the 2024 US Presidential Election. We then used Meta’s tool, CrowdTangle, to query for Instagram posts containing one of these keywords. The CrowdTangle API retrieved data including post date, description, URL, favorite (like) count, and comment count, as well as data of the posting account such as account handle and subscriber count. Our dataset spans April 5, 2024 to August 9, 2024. We then downloaded the image and video content of the curated Instagram posts. This yielded 501,011 images \cite{chang2024visual}. Our keyword list was generated initially from popular election-related hashtags and the names of key election figures. We then used snowball sampling to identify other high-frequency hashtags from the posts within our dataset and expanded our list to include them. In cases where multiple faces appear in an image, we ensure that the emotion analysis results for each face are presented independently. Consequently, the same text may correspond to multiple faces, each with its own emotion analysis result.

\subsection{Text Sentiment Analysis}

Textual data from Instagram posts and stories were analyzed using multiple tools. All posts were selected based on keywords related to the 2024 presidential election, including "Biden," "Trump," "Harris," and others. The data focused on posts and stories shared during the period leading up to the election. To compare visual sentiment with textual sentiment, three sentiment analysis tools were applied: TextBlob \cite{loria2018textblob}, Valence Aware Dictionary and sEntiment Reasoner (VADER) \cite{hutto2014vader}, and a DistilBERT transformer \cite{Sanh2019DistilBERTAD}. TextBlob uses a lexicon-based approach to provide polarity scores on a scale of -1 (the most negative) to +1 (the most positive), and subjectivity scores ranging from 0 (objective) to 1 (subjective) of the given text content. VADER computes normalized, weighted composite scores optimized for social media text on a scale of -1 to +1, with scores above 0.05 classified as positive, scores below -0.05 as negative, and those in between as neutral. The DistilBERT transformer, fine-tuned on the Stanford Sentiment Treebank 2 (SST-2) dataset, achieved an accuracy of 91.3\% on the development set for binary sentiment classification (positive or negative).

\subsection{Image Sentiment Analysis}

\begin{figure}[h]
  \centering
  \includegraphics[width=\linewidth]{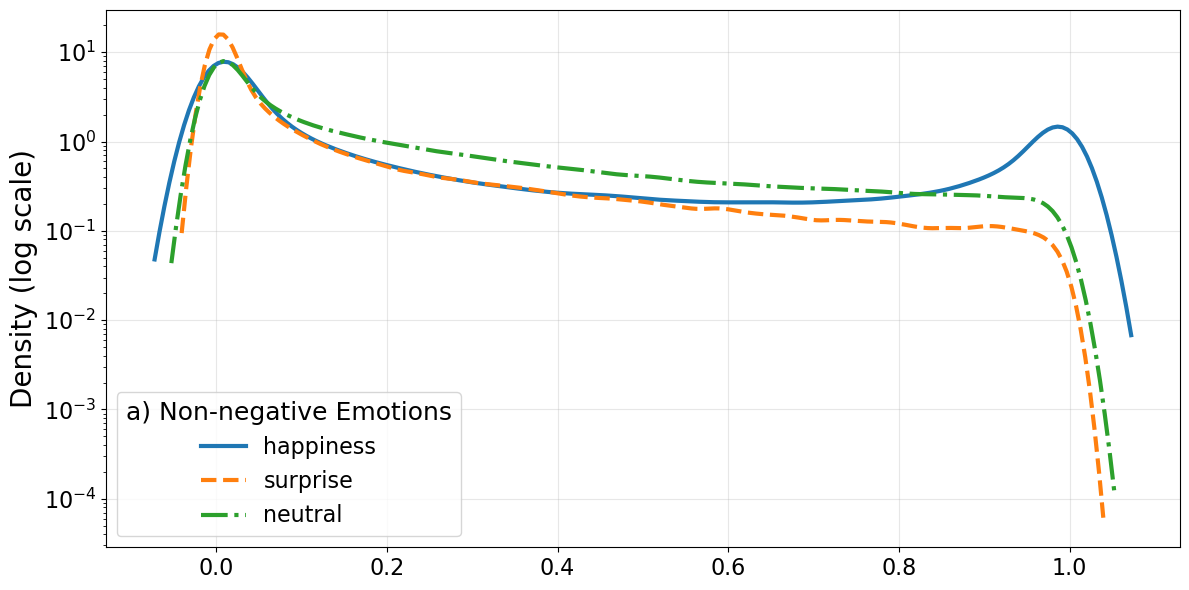}
  \includegraphics[width=\linewidth]{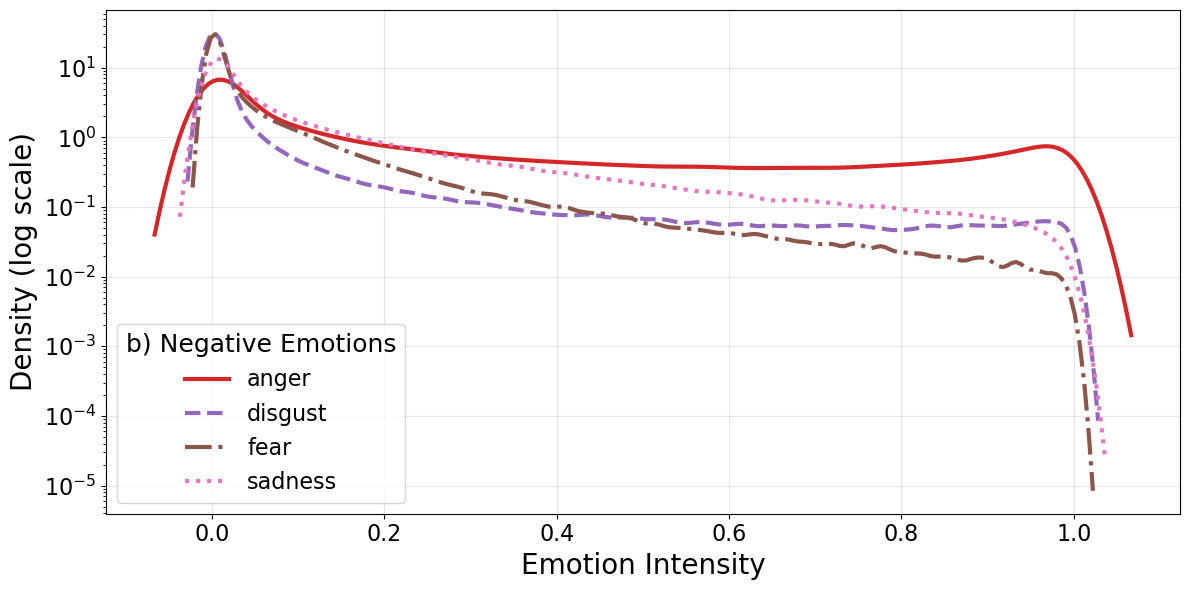}
  \caption{The probability density distributions for a) Non-negative Emotions (Happy, Surprise, and Neutral) and b) Negative Emotions (Anger, Disgust, Fear, and Sadness).}
  \Description{Non-Negative Emotions (Happy, Surprise, and Neutral) Distributions. Negative Emotions (Anger, Disgust, Fear, and Sadness) Distributions.}
  \label{fig:Emotion}
\end{figure}

Images accompanying the posts were analyzed for sentiment using facial expression recognition. Py-Feat is a free, open-source, and easy-to-use tool for working with facial expressions data \cite{cheong2021pyfeat}. In this work, we employed Py-Feat to extract Action Units (AUs), which represent facial muscle movements, and classify emotions into happiness, sadness, anger, fear, surprise, disgust, and neutral. These emotions were quantified with intensity values ranging from 0 to 1, with higher values indicating stronger emotional expressions. Based on the score distribution of non-negative emotions (happy, surprise, and neutral) visualized in Figure ~\ref{fig:Emotion} (a) and negative emotions (anger, disgust, fear, and sadness) in Figure ~\ref{fig:Emotion} (b), a threshold of 0.2 was selected to define the presence of an emotion (e.g., scores above 0.2 indicate the emotion is present). For posts containing multiple faces, the average sentiment scores across all detected faces were computed to represent the overall sentiment of the image content.

Distributions are heavy-tailed and often bimodal, which is reflective of the emotion classification task undertaken by Py-Feat. Importantly, these distributions indicate how facial emotions can appear as a mixture, based on the activating action units.

\subsection{Comparative Analysis}

The study compared text and image sentiments through alignment metrics, such as cross-tabulation of text-based sentiments with image-based emotion categories. We also calculated and compared the weighted average score for each sample, based on the facial emotion scores from Py-Feat and the text sentiment analysis method:

\begin{equation}
\text{Weighted Score of Emotion, } e
= \frac{\sum_{i} \bigl(f_i \times t_i\bigr)}{\sum_{i} f_i}
\end{equation}

Where \( f_i \) represents the facial emotion score for the \( i \)-th face, and \( t_i \) represents the corresponding text sentiment score for the \( i \)-th sample. Since DistilBERT only provides positive and negative labels, we used \(+1\) to represent \textit{Positive} and \(-1\) for \textit{Negative}.

\subsection{Trump Facial Analysis}

To compare partisan image sentiment following major political events, the faces of pro-Democrat and pro-Republican images containing Trump were analyzed before and after he was convicted on May 30, 2024, and before and after his assassination attempt on July 13, 2024. The Instagram dataset was subset to images that contained Trump. Then images posted right before the event and right after the event were analyzed for facial emotions and separated by partisan support. The unit of analysis was faces.

To subset images that contained Trump, we used zero-shot prompting~\cite{puri2019zero} with the \textit{gpt-4o-mini} model to generate an image description and note any people of interest in each image. If either of these labels mentioned Trump, it was included in the subset. The time window of the posts included in the analysis was six days before the event day (5/24-5/29 and 7/7-7/12) for the before and the day of the event plus six days thereafter (5/30-6/5 and 7/13-7/19) for the after. To determine partisan support, we used zero-shot prompting to label each post description as pro-Democrat, against-Democrat, pro-Republican, and against-Republican \cite{chang2024generative}. This method achieved an accuracy of 92\%. Posts labeled as pro-Democrat or against-Republican were classified as Democratic, while posts labeled as pro-Republican or against-Democrat were classified as Republican.

\section{Results} 

\subsection{Visual versus textual sentiment}

\begin{figure}[h]
  \centering
  \includegraphics[width=0.8\linewidth]{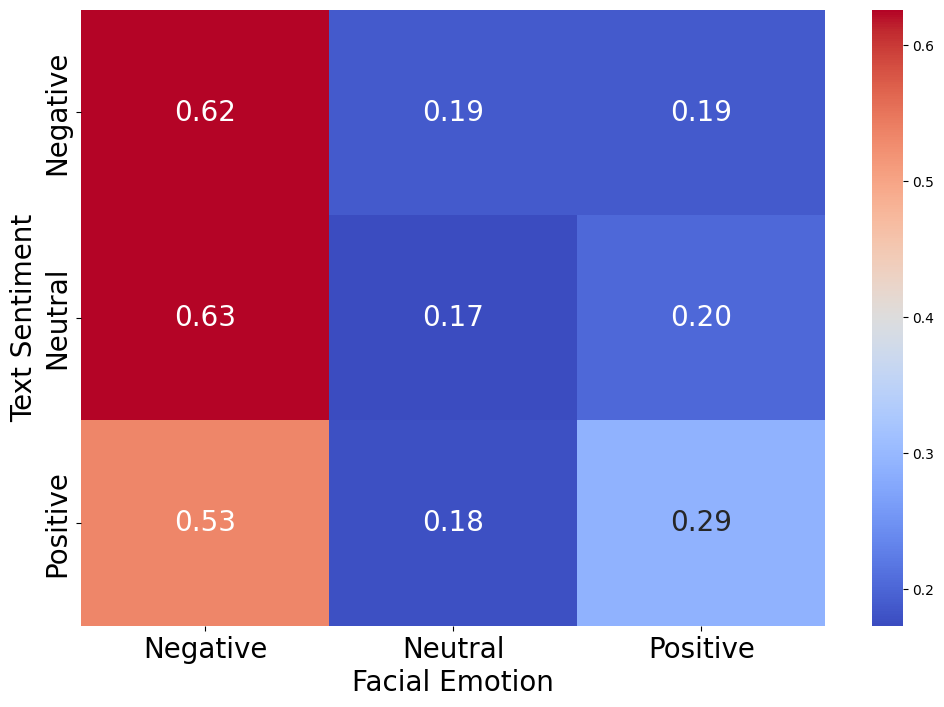}
  \caption{Proportion results of cross-tabulation between text sentiment and facial emotion.}
  \Description{Proportion results of cross-tabulation between text sentiment and facial emotion.}
  \label{fig:cross-tabulation}
\end{figure}

We used the cross-tabulation plot (Figure ~\ref{fig:cross-tabulation}) to demonstrate the relationship between text sentiment analysis and facial detection results. Regardless of whether the sentiment of the text is negative, neutral, or positive, it is predominantly associated with negative facial expressions. For instance, 62\% of negative text sentiment corresponds to negative facial emotions, and even 53\% of positive text sentiment aligns with negative facial expressions, as seen in the proportion analysis. This suggests that the political nature of the topics discussed could cause a stronger tendency toward negative or stern facial emotions. Furthermore, the discrepancy between textual and facial sentiment not only reflects the complexity of emotional expression but also supports our hypothesis, emphasizing the promising potential of integrating facial and textual sentiment analysis for a more comprehensive understanding. 

\begin{figure}[h]
  \centering
  \includegraphics[width=1.05\linewidth]{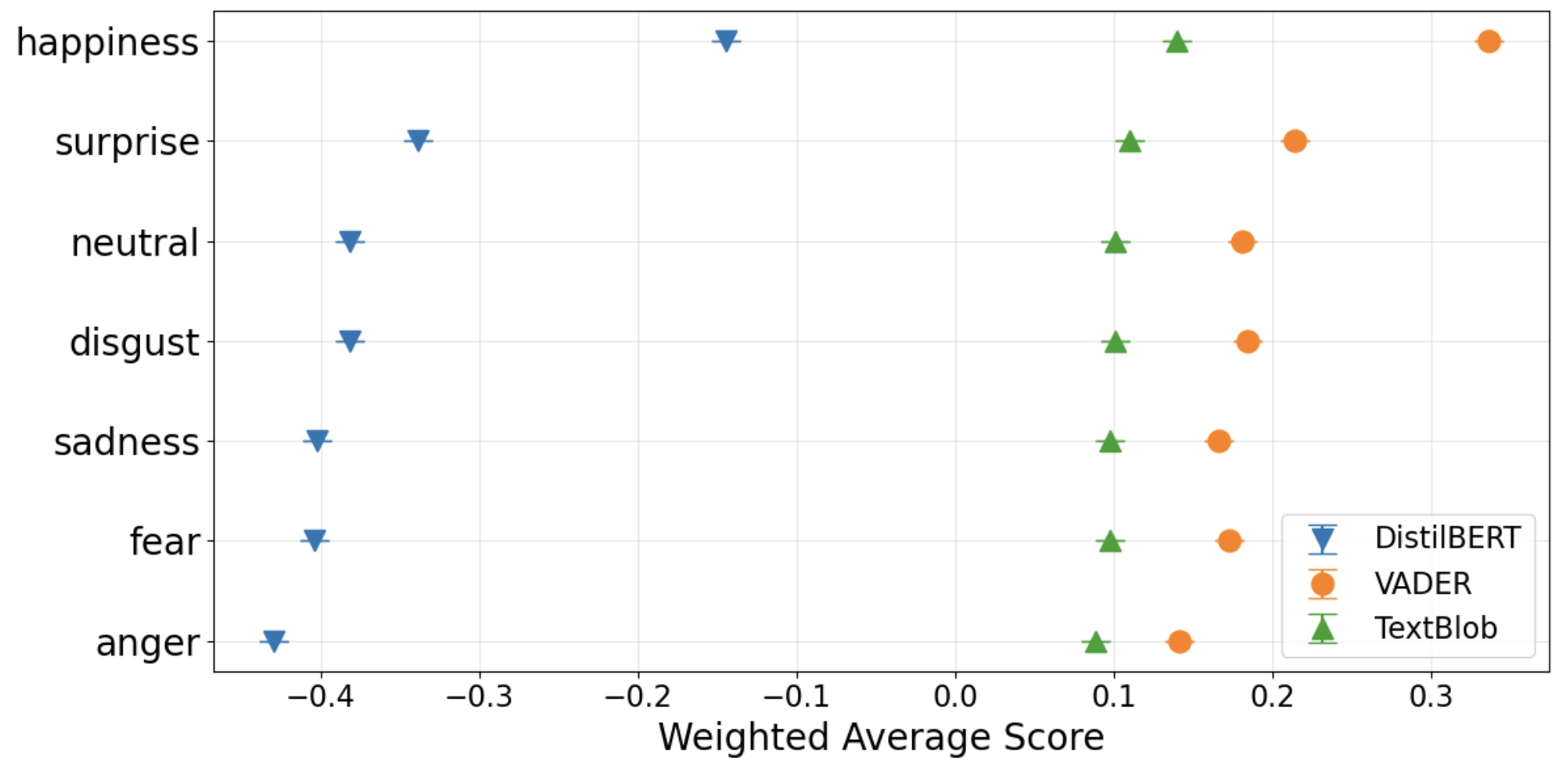}
  \caption{Comparison of Emotion Scores with 95\% Confidence Intervals: DistilBERT, VADER, and TextBlob.}
  \Description{Weighted Average Emotion Scores with Confidence Intervals.}
  \label{fig:weightedAverage}
\end{figure}

 Figure ~\ref{fig:weightedAverage} shows how facial emotions compare with three classic text sentiment methods. For DistilBERT (blue), the weighted score reflects the text sentiment (represented by binary labels) generated by the transformer model on facial emotions. This score is calculated by multiplying the intensity of each facial emotion by the corresponding text sentiment label (with negative sentiment labeled as -1 and positive sentiment as 1), and then averaging the results. The dictionary-based scores--- VADER (orange) and TextBlob (green)---are the mean of the caption averages.
 % The weighted score shows how text sentiment affects changes in facial expressions, with an overall trend that aligns with fluctuations in facial emotions.

Figure ~\ref{fig:weightedAverage} shows that happiness is highly correlated with positive textual sentiment scores across all methods, followed by surprise and neutral expressions. Anger, fear, and sadness consistently yield the lowest scores. Although both scoring methods show similar patterns between text and facial emotions, the resulting values differ. The weighted score of DistillBERT is generally lower because it relies on the binary sentiment labels (negative or positive) generated by the transformer. In contrast, the Vader compound score uses continuous sentiment scores, resulting in higher values. Overall, despite shifts in the scores, all methods reflect a consistent overall trends between text sentiment and facial emotions. Specifically, under the same evaluation criteria, the positive emotion happiness has a relatively higher score compared to other emotions, followed by the more neutral emotions surprise and neutral. 

\subsection{Emotional Shocks and Partisan Divergences}

We demonstrate how facial sentiment can be used for meaningful computational social scientific analyses, by considering how framing of Donald Trump changed before and after two major events during the 2024 USA Presidential Elections. On May 30, 2024, Donald Trump was convicted on all 34 felony counts in his New York hush money trial, becoming the first former U.S. president to be found guilty of felony crimes~\cite{ap2024trump}. On July 13, 2024, Trump survived an assassination attempt during a campaign rally in Butler, Pennsylvania. The attacker, identified as Thomas Matthew Crooks, fired eight shots from a rooftop, grazing Trump's right ear and injuring three rally-goers.

\begin{figure}[h]
  \centering
  \includegraphics[width=1.05\linewidth]{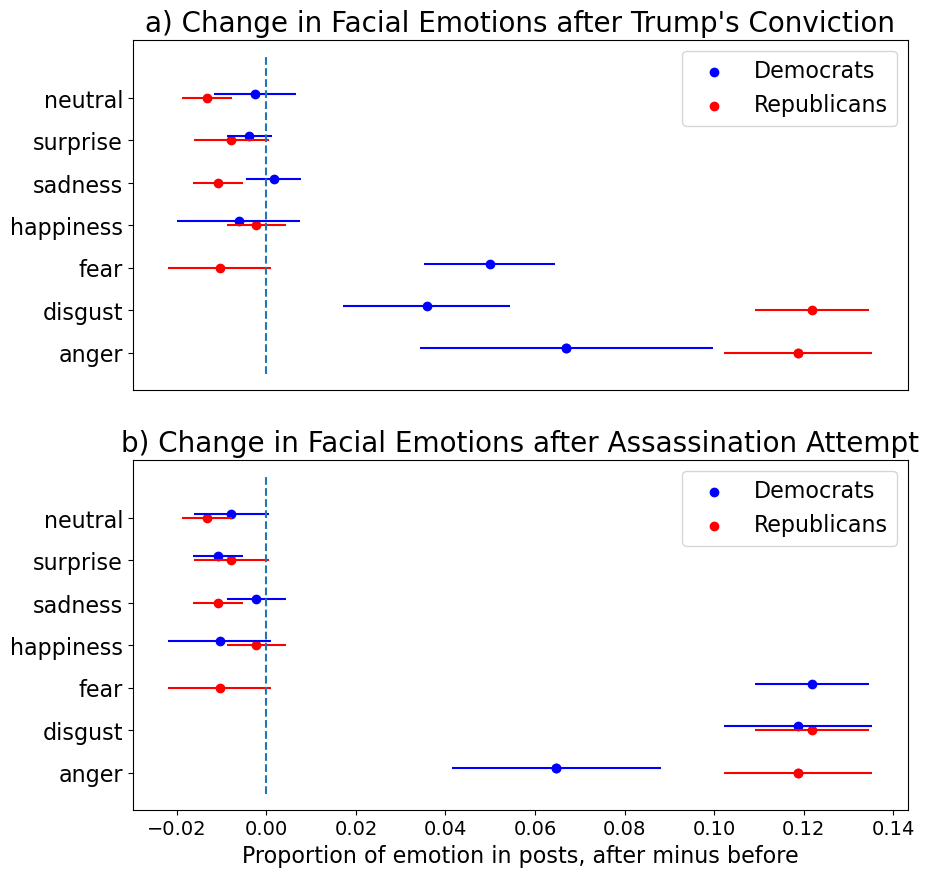}
  \caption{Changes in facial expression before and after a) President Donald Trump's conviction and b) President Trump's assassination attempt.}
  \Description{Forest plot of changes in facial expression of Donald Trump before and after his assassination attempt and conviction.}
  \label{fig:trump}
\end{figure}

Figure~\ref{fig:trump} shows the change in Trump's facial expressions before and after his conviction and assassination attempt. In both cases, we observe up to a 12\% increase in disgust and anger for posts depicting Trump, from both Democrats and Republicans, whereas neutral, surprised, sad, and happy expressions remain roughly the same. This indicates that Trump displays these two negative emotions much more often.

Crucially, there are significant partisan divergences in how Trump is framed. In both cases, Democrats frame Trump with expressions of fear, especially after his assassination attempt. However, Republicans do not frame Trump expressing fear. In other words, the out-group utilizes emotions of weakness for negative campaigning whereas the in-group utilizes anger for strength. Democrats also frame Trump with less disgust than Republicans during his conviction, and less anger overall.

\section{Discussion}

% This paper explores the strengths and limitations of sentiment analysis, highlights the importance of facial emotion recognition as a more direct measure of emotion, and introduces Py-Feat as a critical tool for advancing research in affective computing. By comparing linguistic sentiment analysis with facial affect recognition, we aim to provide a holistic understanding of emotional measurement and offer insights into the most appropriate tools for diverse research contexts.

Our analysis provides evidence that incorporating facial expressions into sentiment analysis offers a valuable complementary perspective to text-based methods. Within the politically charged environment of the 2024 US Presidential Election discourse on Instagram, we observed that facial expressions often diverge from the sentiments expressed in captions. Negative facial expressions were prevalent even in posts with positive or neutral textual sentiment. The results also demonstrate that facial sentiment provides meaningful analytical leverage in identifying patterns across partisan lines. Democrats and Republicans framed Trump differently, with Democrats emphasizing emotions associated with vulnerability (e.g., fear) and Republicans emphasizing anger and disgust---emotions more closely aligned with resilience or defiance. Multimodal sentiment and Py-Feat can uncover partisan framing strategies that would remain hidden if only textual data were considered.

There are several limitations in our work that should be considered. Our work focuses on sentiment analysis in the political domain, specifically applied to facial expressions and text. It does not extend to other popular topics of discussion on social media, such as lifestyle~\cite{djafarova2021instagram} 
, social movements~\cite{chang2022justiceforgeorgefloyd}, 
or public health~\cite{vassey2024scalable,vassey2025worldwide}, which limits the generalizability of our findings. 
Furthermore, our analysis is constrained to Instagram, and the platform’s unique characteristics may introduce biases or limit the applicability of our findings to other platforms with different features and user behaviors. Overall, our results demonstrate the viability of open-source repositories such as Py-Feat, which provide a scalable, multimodal approach to sentiment analysis, opinion mining, and political communication on social media.

\bibliographystyle{ACM-Reference-Format}

\end{document}